# Structural analysis, magnetic and transport properties of $(Ru_{1-x}Co_x)Sr_2GdCu_2O_8$ system

R. Escamilla, A Durán*, and R. Escudero.

Instituto de Investigaciones en Materiales, UNAM, A. Postal 70-360. México D. F.

04510 MÉXICO.

*Centro de Ciencias de la Materia Condensada, UNAM, A. Postal 2681.

Ensenada B. C., 22800, MÉXICO.

**ABSTRACT**

The effects of Co substitution on structural and superconducting properties of $RuSr_2GdCu_2O_8$ compound have been studied. Rietveld refinements of the X-ray diffraction patterns indicate that the cobalt ion progressively replaces ruthenium sites. This replacement induces significant changes on the crystal structure and on the magnetic and superconducting properties. The effects of Co substitution on the superconducting behaviour, and more particularly on the changes induced by the hole doping mechanism, were investigated in $(Ru_{1-x}Co_x)Sr_2GdCu_2O_8$ by a "bond valence sum" analysis with Co content from $x = 0.0$ to $x = 0.2$. The weak ferromagnetic transition at $T_M = 138.2$ K is shifted to lower temperature, and suppressed at higher Co content. From the crystallographic point of view the Ru-O(1)-Cu bond angle, associated to the rotation of the $RuO_6$ octahedra, around the c–axis remain essentially constant when Ru is substituted by Co. Furthermore, increasing Co content has the effect to increase the weak ferromagnetic moment, which may be interpreted as the main responsible for breaking the delicate balance between magnetic and superconducting ordering.

## 1. INTRODUCTION

In $RuSr_2GdCu_2O_8$ (Ru-1212) compound, both ferromagnetic and superconductivity phenomena coexist at the microscopic level [1,2]. The ferromagnetic ordering, appears at a rather high temperature, about 130 – 150 K [3,4], whereas the superconducting transition occurs at low temperature, from about 15 to 45 K. Characteristics of the crystalline structural of this compound can be described based on the high transition temperature compound: $YBa_2Cu_3O_7$ (Y-123). Ru-1212 contains $CuO_2$ planes, which are separated by single oxygen-less Gd planes; $RuO_2$ planes replace the equivalent CuO chains as in the Y-123. In the $RuO_2$ planes, the Ru atoms are six coordinated, and are forming $RuO_6$ octahedra. Additionally, the structure contains SrO planes which are localized between the $CuO_2$ and $RuO_2$ planes. The O(1) oxygen sites are localized in the SrO planes, whereas the oxygen sites O(2) and O(3) are localized in the $CuO_2$ and $RuO_2$ planes, respectively. These planes are connected via the apical oxygen O(1). The $CuO_2$ planes role may be similar to the high $T_C$ cuprates, and directly related to the superconducting formation, whereas $RuO_2$ planes are related to magnetic order. Several experimental studies have been carried out in order to determine the role of cationic substitutions in $RuO_2$ sites [5-7, 9]. For instance, studies of heterovalent substitutions in $Ru_{1-x}M_x$-1212 with (M=$Nb^{V+}$, $Sn^{IV+}$) reveal a reduction of magnetism in the $RuO_2$ planes [8]. Whereas doping Ru sites with $Cu^{II+}$, increases the superconducting transition and reduces the magnetic ordering [9]. Actually, the current understanding of the physical characteristics of this system is that superconductivity and ferromagnetic ordering are originated in the $CuO_2$ and $RuO_2$ planes, respectively. However, the nature of the competition/coexistence of both phenomena so far, requires further understanding. In this context we are reporting studies on the effect of cobalt substitution in Ru sites. The study emphasizes on the changes in the crystalline structure and effects on the superconducting and magnetic properties.

## 2. EXPERIMENTAL

Polycrystalline samples of $(Ru_{1-x}Co_x)Sr_2GdCu_2O_8$ (x = 0, 0.025, 0.05, 0.075, 0.1, 0.2) were synthesized by solid state reaction of oxides: $RuO_2$ (99%), CoO (99.999%), $Gd_2O_3$ (99.9%), CuO (99.99% and $SrCO_3$ (98+%). After calcinations in air at 900 $^0$C, the samples were grounded, pressed into pellets and annealed in oxygen at 1000 $^0$C. Phase identification was performed using a X-ray diffractometer, Siemens D5000 with Cu-$K_\alpha$ radiation and Ni filter. Intensities were measured in steps of 0.02 degrees for 14 seconds in the 2θ range from 5 to 120, at room temperature. Crystallographic phases were identified by comparison with X-ray patterns in the JCPDS database. The crystalline structure was refined with the program Rietica [10] (Rietveld program for quantitative phase analysis of polycrystalline mixtures with multi-phase capability). The Bond Valence Sum program was used to distinguish the oxidation states of metals Ru and Cu [11]. The superconducting transition temperatures were determined with a closed-cycle helium refrigerator by measuring the resistance vs. temperature characteristic by the standard four-probe technique, from 250 K to 14 K. Measurements of ac susceptibility, dc susceptibility, and magnetization versus magnetic field were carried out using a superconducting Quantum Interference Device Magnetometer, (MPMS Quantum Design), from 2 to 300 K and applied field up to ± 40 kOe.

**3. RESULTS AND DISCUSSION**

Figure 1a shows X-ray diffraction patterns for the $Ru_{1-x}Co_x$-1212 samples. The structural analysis indicate that all samples correspond to the Ru-1212 structure, with negligible content of impurities. In figure 1b we also show an amplification of the region on the X-Ray data where impurities were detected at the 2θ position, these correspond to $Sr_3(Ru,Cu)O_7$ (ICDD n° 51-0307), and $SrRuO_3$ (ICDD n° 28-1250). We can note that those are negligible, inclusive at the maximum Co concentration. In the refinement process we took into account the presence of secondary phases and the substitution of Co ions in Ru and Cu sites. Fig. 2 shows an example of a fitted patterns for the un-doped sample. More detailed of the structural characteristics are listed in Table 1, the first three rows show the trend of change of the lattice parameters at room temperature with increase cobalt content. It is observed that as x is increased, the c-axis parameter decreases, mainly as a result of the decrease of the Ru/Co-O(1) average bond length. The a-axis increases slightly, and as consequence of those changes the unit-cell volume decreases slightly. This result may be explained considering the co-ordination numbers and the ionic radii of the $Co^{III+}$, $Cu^{II+}$ and $Ru^{V+}$ ions. It is noted that the $Co^{III+}$ radius with six co-ordination number changes from 0.545 to 0.61 Å for low spin and high spin configuration, respectively. Whereas $Cu^{II+}$ with a size of 0.65 Å in five co-ordination number, and $Ru^{V+}$ in six co-ordination number have a size of 0.565 Å [12]. From these values, it is clear that the decrease of the unit cell volume can be related to the substitution of $Co^{III+}$ in low spin configuration state into the $Ru^{V+}$ sites. This behaviour is consistent with the observed changes in $Y_1Ba_2Cu_3O_7$, when $Co^{III+}$ is substituted in low spin configuration into the CuO chains [13].

From results of our structural refinement, we can conclude that cobalt atoms majority occupy the Ru sites, which give place to substantial changes on the Ru/Co– O(1) bond length, rather that in the Ru/Co – O(3) bond length. As a consequence, an increase in the octahedral distortion ($\Delta_{oct}$) is observed; from 0.184 Å (x = 0.0) to 0.287 Å (x = 0.2), see Table 2. Furthermore, the Ru – O(1) – Cu bond angle associated to the rotation of $RuO_6$ octahedra around the c–axis, remains essentially constant. In contrast, the Cu–O(2)–Cu buckling angle changes from 169.2° (x = 0.0) to 167.99° (x = 0.2).

On the other hand, the bond valence sum (BVS) shows that as x is increased, the valence of the Ru increases whereas the Cu decreases, see Table 2. For un doped, samples the calculated Ru valence was 4.69. This result is in agreement with the obtained by NMR and XANES studies; which indicate mixed ionic states about 40% $Ru^{IV+}$ and 60% $Ru^{V+}$ [14, 15]. Thus, the main effect of Co substitution is a strong increase of the BVS at the Ru site, accompanied by an important decrease of the BVS at Cu site. In table 1 we show these results.

Temperature dependence of normalized resistance at zero external field for all samples is shown in Fig. 3. For the sample with x = 0, the resistance shows a metallic-like behaviour with a superconductivity onset at about 57 K, and reaching zero resistance at 33 K. Contrarily, doped samples do not display superconductivity below 10 K. At composition with x=0.025, the normal state resistance starts to increase at high temperature with a downturn at low temperature, while for higher compositions x > 0.025 a semiconducting-like behaviour is observed. In the inset of Fig. 3 we show the real part of the ac susceptibility data, for the un doped sample. A small upward at about 138.2 K marks the onset of the spontaneous magnetisation, indicating the weak ferromagnetic transitions ($T_M$). At 35 K it is noted a continuous decreasing of the ac susceptibility indicating the superconducting transition, but without diamagnetic signal. This fact might indicate either microstructural effects connected with sample granularity (macroscopic screening currents) [16] or competing effects between magnetism and superconducting ordering as have been observed in rare earth intermetallic compounds [17]. These ac data show that our samples are structurally comparable to other samples produced and reported by other laboratories [3, 4].

In order to explain the $T_C$ degradation in ours samples, we determine the amount of charge ($p$) transferred between the $CuO_2$ and $RuO_2$ planes. We took into account the simple model of valence (Ru:5-2$p$, Cu:2+$p$) [6]. The $p$ value comes from the structural data using BVS that make use of the sensitivity of Cu-O bond lengths at the hole concentration [18]. Our calculations show that with increasing x, $p$ changes from 0.155 to 0.1, and from 0.05 to -0.4 for Ru and Cu respectively, see Table 2. Moreover, Jorgensen *et al* [19] propose that the highest $T_C$ is achieved in structures with flat and square $CuO_2$ planes and long apical Cu–O bond length. This fact rest on the assumption that the buckling of $CuO_2$ planes, and the shortening of the apical Cu-O bond, localize holes. Ours results reveal that as x is increased, the buckling of the $CuO_2$ plane is increased, and the apical Cu-O bond length decreases. Therefore, the $T_C$ degradation may be associated with the increase of the buckling of the $CuO_2$ planes and to the decreases of the apical Cu-O bond length, due to changes of charge ($p$) in the $CuO_2$ and $RuO_2$ planes.

We performed measurements of dc susceptibility as a function of temperature at low magnetic field (15 Oe), in zero field cooling (ZFC) and field cooling (FC) conditions, from 300 to 2 K as shown in figure 4. The Meissner effect is not observed there. The inset of Fig. 4 shows the effect of Co substitution on the magnetic characteristics for all samples. The ferromagnetic $T_M$ and the antiferromagnetic $T_N$ transitions are clearly noted with help of the susceptibility derivative data ($d\chi/dT$). The results reveal that $T_M$ decreases as Co is increased, and disappears at a concentration of x = 0.200. Moreover, in the range of compositions for x < 0.10 we observed a clear splitting of the ZFC and FC measurements close to $T_M$, that disappears for x = 0.200 (inset presents only ZFC measurement). It is important to note that not only the Co concentration has effect on $T_M$, but also the external magnetic field; as an example of this behaviour, we show in Fig. 5 measurements of $\chi$ - T for the Cobalt composition x = 0.10. These measurements were performed at ZFC and FC. The splitting of the data is clearly observed at low fields with two branches (note that the high susceptibility branch is FC). However, as the field is higher the splitting is small and disappears at 10 and 20 kOe. The extended extracted data for measurements in other Co compositions are illustrated in Fig. 6a. There, it seems that as the Co concentration is increased, the magnetic transition $T_M$ is shifted to low temperature, but depending also on the applied field. Furthermore, increasing the magnetic field, shifts $T_M$ to high temperature (this also depend on the Co concentration), for example, at 1 kOe $T_M$ is about 133 K when x = 0.025, but when the field is 30 kOe, $T_M$ is shifted to about 155 K. $T_M$ disappears for the composition with x = 0.10 when the field is 30 kOe. Thus, both the Co concentration and the magnetic field may destroy $T_M$. On the other hand, in Fig. 6b we show that $T_N$ is affected notably by the intensity of the magnetic field, with no effect in Co concentration. In this figure is observed that $T_N$ is shifted from 8 K at 1 kOe to about 2.6 K at 30 kOe. This behaviour is quite typical of layered structures; the external magnetic field tends to saturate the paramagnetic moment of the Gd sub lattice [20]. Thus, the susceptibility results suggest that the dominant magnetic character is antiferromagnetic order (AFM) at low temperatures, and canted ferromagnetism at high temperature since the transverse components of spins differ from one crystallographic site to another and the equilibrium angles depend on both the external magnetic field and dopant content. These experimental evidences have been well documented in doped manganites, cuprates and intermetallic compounds [20-22]. In order to understand this susceptibility behaviour we can assume that two independent contributions exist; one for each Gd and Ru sub lattices. The Ru sub lattice orders as a weak ferromagnet at Tm, while the Gd one, remains paramagnetic and orders antiferromagnetically at lower temperature. To separate the two contributions and to observe the behaviour as a function of Co content, we estimated, at high-temperature, the Ru moments by fitting the susceptibility data considering two contributions, i.e. $\chi^{-1} = [(C/(T-\Theta))_{Gd} + (C/(T-\Theta) + \chi_0)_{Ru}]^{-1}$. In the first term the Curie constant $C_{Gd}$ was taken as the theoretical value of 7.9 emu-K/mol and the Curie temperature as $\Theta_{Gd}$ = -9 K. This value for $\Theta_{Gd}$ was used by Butera et al. [23,24], and was obtained performing EPR measurements. In the second term we included the Curie constant $C_{Ru}$, the Curie temperature ($\Theta_{Ru}$) and the susceptibility $\chi_0$ which is temperature independent; because the non-linearity of the inverse susceptibility data in the measured range (170 to 300 K) as seen in Fig 7. In this fitting process both $C_{Gd}$ and $\Theta_{Gd}$ were kept fixed in order to observe the Ru effective moment behaviour, these are plotted in inset of Fig. 7.

A continuous increase of the effective moments $\mu_{eff}$ is seen from ~1.06 $\mu_B$, for x = 0 to 2.17 $\mu_B$ at x = 0.075. Increasing Co atoms, affects $\mu_B$, reducing it to a value of about 1.28 $\mu_B$. On the other hand, the Curie temperature, $\Theta_{Ru}$ changes from 153.9 ±0.5 to 112 ±2.5 degrees, for 0.025 <x<0.10 and then increases again from x > 0.10 to a value about 157.7 ±1.2 degrees. To have a better understanding of this magnetic behaviour, we performed magnetisation versus applied magnetic field (M-H) measurements at T = 2 K (see Fig 8). The hysteresis loops at low field (main panel) is attributed to the weak magnetism of the Ru/Co sub lattice, whereas the dependence of the magnetization at high field region (inset b) to the antiferromagnetic order of Gd ion. A trend to saturation is observed with value at about 7.8 $\mu_B/f.u$, which correspond to the Gd moment, as obtained by neutron diffraction studies (~7.0 µB ) [25]. This experimental value confirms the two expected contributions: The Gd moment, plus 0.8 $\mu_B/f.u.$ due to ruthenium sub lattice at low spin state (1$\mu_B$ for 3d$^3$, S=1/2). On the other hand, the behaviour of the remanent magnetization ($M_R$) associated to the Ru sub lattice as dependence of Co content is plotted in inset A). $M_R$ is increased from 0.14 $\mu_B/f.u.$ for x = 0 to 0.49 $\mu_B$/f.u. for x = 0.05 and then decreases to 0.05 $\mu_B$/f.u. at a concentration about x = 0.2. The small value of $M_R$ for x = 0 is consistent with the analysis of neutron diffraction measurements by Lynn et al [25], where an upper limit of about 0.1 $\mu_B$ was obtained for the ferromagnetic peaks below of $T_M$. Again, it is worthwhile to note that $M_R$ as well as $\mu_{eff}$ extracted from Ru sub lattice (inset of Fig. 7) are increased as Co is introduced in the Ru sub lattice (for x < 0.10). This behaviour is opposed to Sn and Nb as in Eu-1212 [26]. However, these characteristics are still more intriguing in the Ce,Eu-1222 [27] and Gd-1212 [28] compounds doped with iron where a decreasing of the magnetic moment in the Ru sub lattice was observed. It is possibly that this increasing of the magnetic moment in the Ru/Co site may be due to an enhancement of the antisymmetric exchange, causing the effect that spins be canted in the ab-plane as it has been argued in several intermetallic and ruthenocuprate compounds [20,22,29,30], where a delicate balance among the subtle variation in composition, magnetic structure, and superconducting state may exist. Furthermore, recent studies [31] have shown that complete substitution; Ru by Co, gives a paramagnetic material which is accomplished by depleting oxygen in the structure. The drastic reduction of magnetic moment for x > 0.1 accompanied with the strong octahedral distortion observed here could be related to this fact.

## 4. CONCLUSIONS

In summary, we have presented a detailed crystallographic and magnetic study of the $(Ru_{1-x}Co_x)Sr_2GdCu_2O_8$ system by X-ray diffraction at room temperature and magnetisation measurements. X-ray diffraction results indicate that Co ions occupy the Ru sites. This replacement causes significant changes in the Ru - O(1) bond length inducing an increase in the octahedral distortion. Contrarily to previous report, we found that the bond angle associated with the rotation around the c-axis remains essentially constant. From the BVS analysis, we demonstrate that as x is increased, the amount of charge (*p*) decreases in the $CuO_2$ and $RuO_2$ planes, this could explain the observed $T_C$ reduction with Co content. Finally, we found from magnetic measurements that a gradual introduction of Co-ions into the structure, enhance the magnetic moment, $\mu_{eff}$, and $M_R$ via the ferromagnetic component. Thus we believe that this increase of the magnetic moment might be responsible for the delicate balance between weak ferromagnetism and superconductivity via pair breaking or trapping holes in the $CuO_2$ planes.

**Acknowledgments**

Economical support is acknowledged to UC-MEXUS. M. Sainz and F. Silvar for technical support and D. Bucio for assistance in sample preparation

**FIGURE CAPTION**

Fig. 1a X - ray diffraction patterns for the $Ru_{1-x}Co_x$-1212 samples. The symbols (+ and *) represent the position of the impurities of $SrRuO_3$, and $Sr_3(Ru,Cu)O_7$.

Fig. 1b Amplification of the region where impurities are detected in the main pattern of Fig. 1a, these are in $2\theta$ region from 31-35 degrees, and from 45-48 degrees, and correspond to $SrRuO_3$, and $Sr_3(Ru,Cu)O_7$, as marked in this figure.

Fig. 2 Rietveld refinement on the X-ray diffraction pattern for the x = 0.0 sample. Experimental spectrum (dots), calculated pattern (continuous line), difference (middle line) and the calculated peaks positions (bottom).

Fig. 3 Normalized resistances as a function of temperature in $(Ru_{1-x}Co_x)$-1212 for different Co concentrations as indicated in the figure. Inset shows ac susceptibility for x = 0.0 measured at 730 Hz and field amplitude of 3.0 Oe.

Fig. 4 dc susceptibility measurements as a function of temperature for $(Ru_{1-x}Co_x)$-1212. We show the ZFC and FC modes with an external magnetic field of 15 Oe. The inset shows $d\chi/dT$ vs T for $0<x<0.2$ composition.

Fig. 5 $\chi$ vs T measurements (ZFC and FC modes) at several external magnetic fields for $(Ru_{1-x}Co_x)$-1212 for the cobalt content x=0.10

Fig. 6 a) Behaviour of $T_M$ as a function of Co content at several applied magnetic fields  b) Antiferromagnetic transition $T_N$ as a function of applied magnetic field. $T_N$ is independent of Co content.

Fig. 7. Temperature-dependent of the inverse molar susceptibility $\chi^{-1}$ for $(Ru_{1-x}Co_x)$-1212 with $0.025 > x > 0.2$ at  1 kOe. Inset shows the effective moments $\mu_{eff}$ and the Curie-Weiss temperature for Ru sub lattice as a function of Co content (the line is guided to the eye).

Fig. 8. Hysteresis loops in the low field region for $(Ru_{1-x}Co_x)$-1212 with x = 0 (full square), 0.05 (open triangles), 0.10 ( full triangles) and 0.20 (solid line). Inset a) shows the remanent moment ($M_R$) as a function of Co content (the solid line is guided to the eye). Inset b) shows magnetisation at high field  40 kOe for x = 0 (full square), x = 0.075 (open circle) and x = 0.2 (solid line).

**TABLE CAPTION**

Table 1 Structural Parameters for $Ru_{1-x}Co_x$ -1212 at 295 K.

Table 2 Bond lengths (Å), bond angles (deg) and octahedral distortion ($\Delta_{oct}$) for $Ru_{1-x}Co_x$-1212.

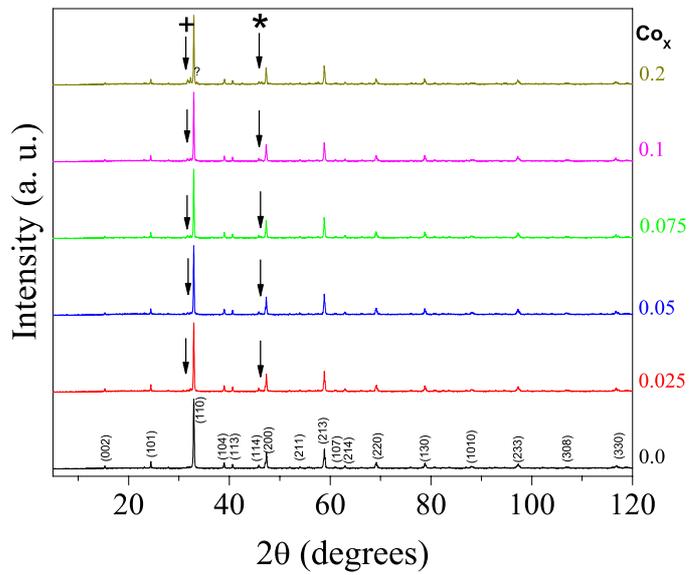

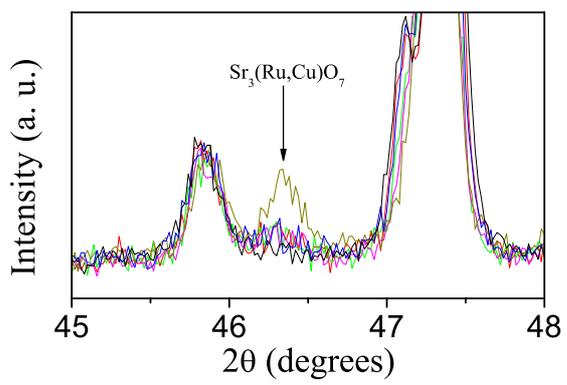
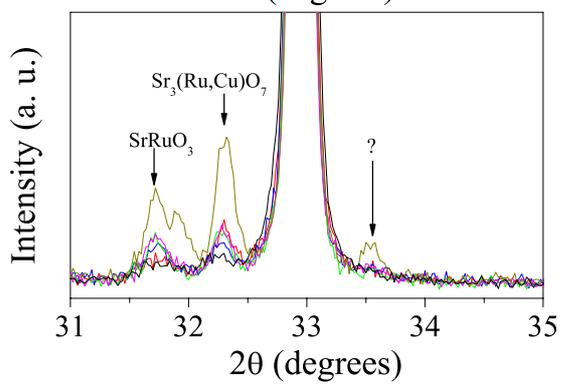

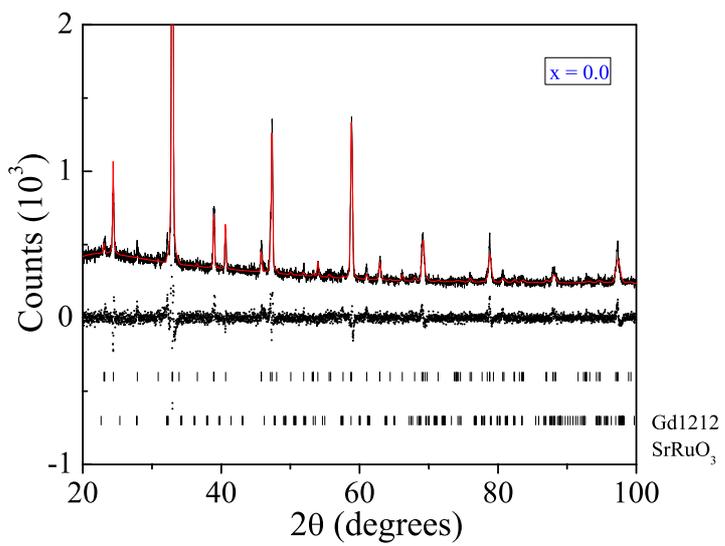

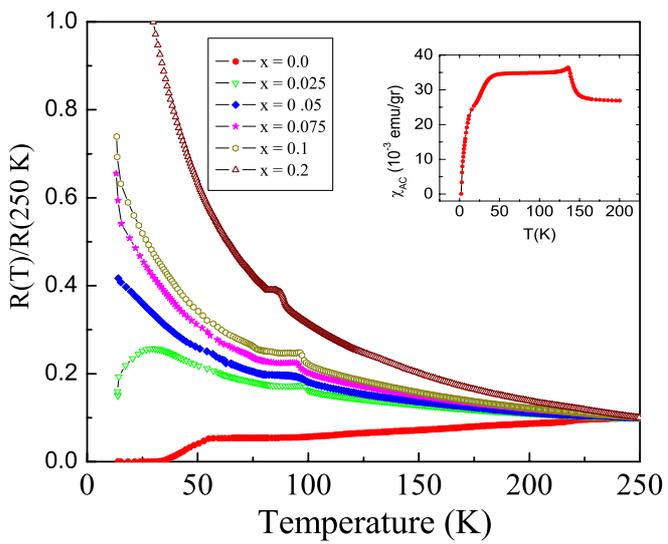

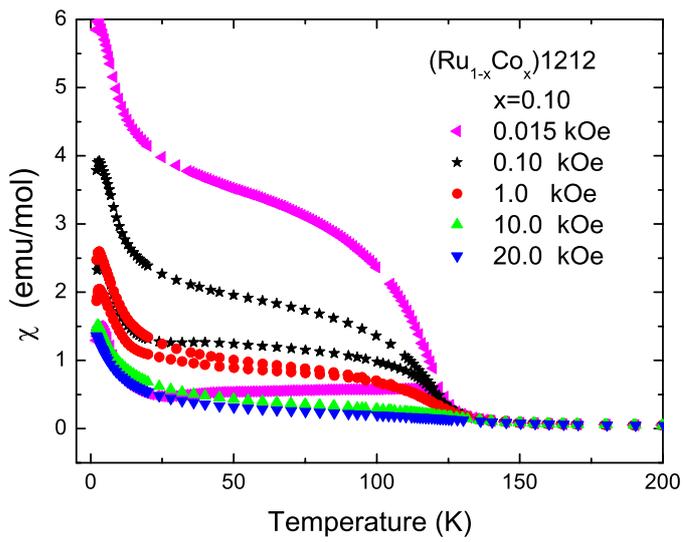

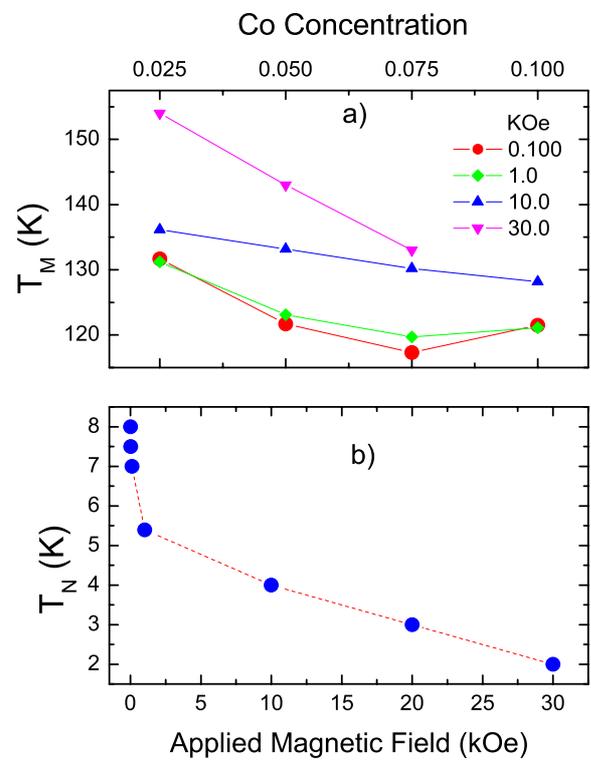

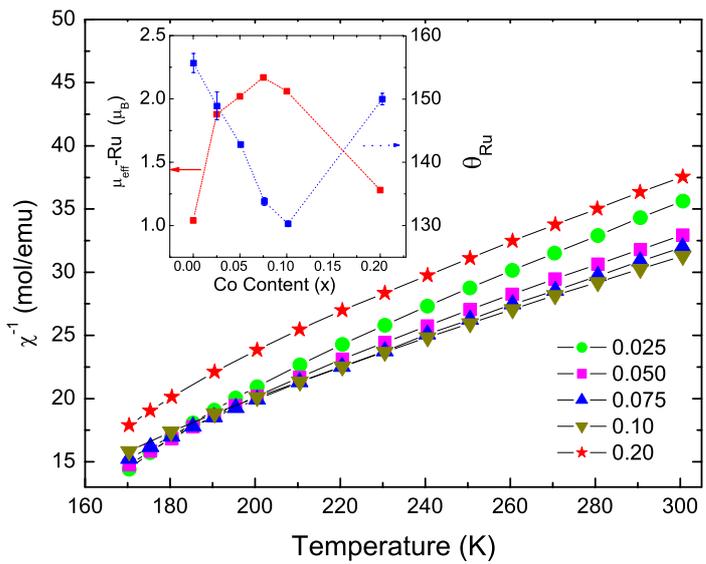

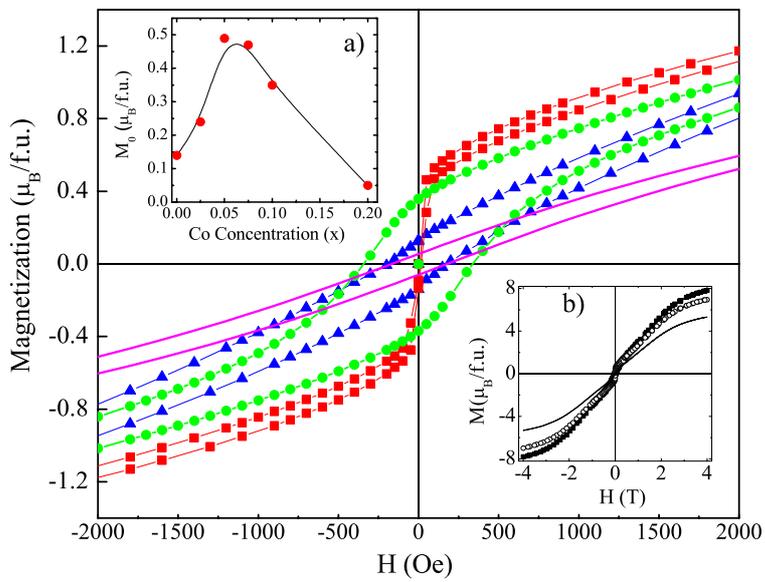

# TABLE 1

| ( x =) | | 0.0 | 0.025 | 0.05 | 0.075 | 0.1 | 0.2 |
|---|---|---|---|---|---|---|---|
| | a(Å) | 3.8367(3) | 3.8374(2) | 3.8382(2) | 3.8388(4) | 3.8393(4) | 3.8408(4) |
| | c(Å) | 11.568(3) | 11.5622(2) | 11.5584(3) | 11.5544(3) | 11.5505(4) | 11.5417(4) |
| | V(Å$^3$) | 170.29 | 170.26 | 170.28 | 170.27 | 170.25 | 170.25 |
| Ru/Co | $S_{ij}$ (v.u.) | **4.69** | **4.71** | **4.72** | **4.73** | **4.79** | **4.80** |
| | p | **0.155** | **0.145** | **0.14** | **0.135** | **0.105** | **0.1** |
| | B(Å$^2$) | 0.82(9) | 0.86(8) | 1.39(8) | 0.82(8) | 0.74(8) | 1.5(1) |
| | N | - | 0.02(1) | 0.04(2) | 0.07(2) | 0.09(1) | 0.18(1) |
| Gd | B(Å$^2$) | 2.06(7) | 1.72(6) | 1.52(6) | 2.09(6) | 1.53(5) | 2.09(9) |
| Sr | z | 0.3067(4) | 0.3067(4) | 0.3066(2) | 0.3066(2) | 0.3066(2) | 0.3066(2) |
| | B(Å$^2$) | 0.84(7) | 0.50(5) | 0.99(7) | 0.70(5) | 0.54(5) | 0.25(7) |
| Cu | z | 0.1452(2) | 0.1455(2) | 0.1456(3) | 0.1457(3) | 0.1457(3) | 0.1470(3) |
| | $S_{ij}$ (v.u.) | **2.05** | **2.04** | **2.04** | **1.62** | **1.61** | **1.60** |
| | p | **0.05** | **0.04** | **0.04** | **-0.38** | **-0.39** | **-0.4** |
| | B(Å$^2$) | 1.04(8) | 1.40(7) | 0.70(8) | 0.43(23) | 0.98(6) | 1.2(1) |
| O(1) | x | 0.0390(1) | 0.0389(2) | 0.0390(1) | 0.0390(1) | 0.0390(2) | 0.0390(2) |
| | z | 0.3335(3) | 0.3337(3) | 0.3338(4) | 0.3339(4) | 0.3347(4) | 0.3347(4) |
| | B(Å$^2$) | 5.8(8) | 5.4(7) | 3.9(6) | 7.2(7) | 6.9(7) | 8.5(1.3) |
| O(2) | z | 0.1295(1) | 0.1297(4) | 0.1295(3) | 0.1295(3) | 0.1295(4) | 0.1295(4) |
| | B(Å$^2$) | 2.9(4) | 2.1(3) | 0.7(2) | 1.0(3) | 2.0(3) | 1.9(5) |
| O(3) | x | 0.1139(1) | 0.1140(1) | 0.1141(2) | 0.1140(1) | 0.1140(1) | 0.1140(2) |
| | B(Å$^2$) | 5.8(6) | 5.8(5) | 5.7(4) | 5.6(4) | 5.1(5) | 5.1(8) |
| Gd1212 | | 94.91(2) | 95.83(1) | 95.43(1) | 93.28(2) | 86.16(3) | 83.40(6) |
| %SrRuO$_3$ | | 2.1(1) | 3.8(1) | 2.70(8) | 5.1(1) | 3.1(1) | 8.0(1) |
| %Sr$_3$Ru$_2$O$_7$ | | 2.9(4) | 0.3(2) | 1.9(2) | 1.6(3) | 3.3(3) | 6.7(8) |
| | $R_p$ (%) | 8.8 | 6.0 | 5.9 | 6.0 | 5.9 | 6.5 |
| | $R_{wp}$(%) | 11.5 | 8.3 | 8.2 | 8.7 | 8.0 | 9.1 |
| | $R_{exp}$(%) | 9.2 | 3.7 | 3.6 | 3.6 | 3.6 | 3.7 |
| | $\chi^2$(%) | 1.2 | 2.2 | 2.3 | 2.4 | 2.2 | 2.5 |

Note. Space group: *P4/mmm* (# 123). $S_{ij}$ (v.u.) is the bond valences sum, *p* is amount of charge transferred between the CuO$_2$ and RuO$_2$ plane. N is the cobalt occupancy factor, Atomic positions: Ru: 1b (0, 0, ½); Gd: 1c (1/2, ½, 0); Sr : 2h (1/2, 1/2, z); Cu: 2g (0, 0, z); 2O(1) in 8s (x,0,z) × 1/4, 4 O(2) in 4i (0,1/2,z), and 2O(3) in 4o (x,1/2,1/2) × 1/2 position.
% of impurity in the phase

**TABLE 2**

| $x =$ | 0.0 | 0.025 | 0.05 | 0.075 | 0.1 | 0.2 |
|---|---|---|---|---|---|---|
| Ru- O(1): 2 | 1.931(2) | 1.931(2) | 1.928(2) | 1.926(2) | 1.924(2) | 1.914(2) |
| Ru – O(3): 4 | 1.967(2) | 1.968(3) | 1.968(1) | 1.969(1) | 1.969(3) | 1.970(1) |
| ⟨Ru – O⟩$_{average}$ | 1.956 | 1.956 | 1.955 | 1.955 | 1.954 | 1.951 |
| $\Delta_{oct}$ | 0.184 | 0.189 | 0.205 | 0.220 | 0.230 | 0.287 |
| Cu - O(1) | 2.182(3) | 2.182(2) | 2.180(2) | 2.180(2) | 2.179(2) | 2.171(2) |
| Cu - O(2) | 1.927(2) | 1.927(1) | 1.928(2) | 1.928(1) | 1.929(1) | 1.931(1) |
| Ru-O(3)-Ru | 154.3(2) | 154.3(3) | 154.3(1) | 154.3(1) | 154.3(1) | 154.3(2) |
| Cu-O(2)-Cu | 169.2(1) | 169.1(2) | 168.9(2) | 168.9(3) | 168.0(2) | 168.0(1) |
| Ru-O(1)-Cu | 171.6(2) | 171.6(2) | 171.6(2) | 171.6(2) | 171.6(2) | 171.6(2) |

$\Delta_{oct}$: **octahedral distortion**